\documentclass[doublecol,a4paper]{epl2}

\usepackage{makeidx}
\usepackage{amssymb}
\usepackage{amssymb,amsmath}
\usepackage{graphicx}
\usepackage{graphics}
\usepackage{simplewick}

\newcommand{\PathToFigures}{}

\title{Effect of disorder on 2D topological merging transition \\
from a Dirac semi-metal to a normal insulator}

\shorttitle{Effect of disorder on 2D topological merging transition} 

\author{David Carpentier  \and  Andrei A. Fedorenko \and Edmond Orignac}
\shortauthor{D. Carpentier  \etal}
\institute{ CNRS UMR5672 - Laboratoire de Physique de l'Ecole Normale
Sup{\'e}rieure de Lyon,\\ 46, All{\'e}e d'Italie, 69007 Lyon, France\\
}

\abstract{
We study the influence of disorder  on  the topological transition from
a two-dimensional Dirac semi-metal to an insulating state. This transition is
described as a continuous merging of two Dirac points leading to a semi-Dirac
spectrum at the critical point. The latter is characterized by a dispersion
relation linear in one direction and quadratic in the orthogonal one.
Using the self-consistent Born approximation and renormalization
group we calculate the density of states above, below and in the vicinity of the
transition in the presence of different types of disorder.
Beyond the expected disorder smearing of the transition we find an
intermediate disordered semi-Dirac phase. On one side this phase is separated
from the insulating state by a continuous transition while on the other side
it evolves through a crossover to the disordered Dirac phase.
}

\begin{document}

\maketitle

\section{Introduction}

In the recent years  there has been a growing interest for materials whose
low energy degrees of freedom are described by two dimensional (2D)  Dirac fermions.
For instance, 2D Dirac points have been found in the energy spectrum of
graphene~\cite{geim07,castroneto09},
3D topological insulators \cite{fu07,moore07} such as the materials in the
Bi$_{2}$Se$_{3}$ family\cite{hsieh08} and strained HgTe~\cite{hancock11},
quasi-2D organic conductor $\alpha$-(BEDT-TTF)$_{2}$ I$_{3}$ under
pressure \cite{katayama06,kobayashi07,goerbig08,nishine10}
and very tunable molecular system assembled by atomic manipulation of
carbon monoxide molecules on a copper surface~\cite{gomes12}.
Each Dirac point carries a topological charge which is robust against
perturbations~\cite{wunsch08}. This topological charge is
related to the chirality of the electronic states, but can also be viewed
as a Berry monopole. The Berry phase associated with any closed path in
reciprocal space is determined
by the total topological charge of the Dirac points enclosed by the path.
Moreover, in the case of two-dimensional models, excluding the surface states,
according to the fermion doubling theorem the Dirac cones emerge by pairs
of opposite topological charges. In this case, a
topological transition  towards an insulator is possible
which consists of a merging of the Dirac points of opposite
topological charges~\cite{montambaux09-epjb}.
Moving the Dirac points in the Brillouin
zone up to their fusion requires perturbations of the underlying
microscopic Hamiltonian. This can be achieved, for example, by mechanical
stretching or depositing adatoms that could give rise to
a hopping anisotropy or enhancement of higher-order hopping amplitudes.
Exactly at the topological transition triggered by the merging of two Dirac points,
a very peculiar {\it semi-Dirac dispersion} relation occurs, which is linear in one
direction while being quadratic in the orthogonal
one~\cite{montambaux09-epjb,dietl08,volovik01,lim12,delplace10}.
Recently such a merging of tunable Dirac points has been
realized using a degenerate Fermi gas trapped in a 2D honeycomb
optical lattice~\cite{tarruell12}, in photonic graphene~\cite{rechtsman12}
and in microwave experiments~\cite{bellec13}.
Moving and merging of two Dirac
points in the organic conductor $\alpha$-(BEDT-TTF)$_{2}$ I$_{3}$ has
also been discussed in~\cite{katayama06,kobayashi11,suzumura13}.

For 2D Dirac semi-metals, both the density of states (DOS) near the Dirac point
and the transport properties may be strongly affected by the presence of disorder.
Its effect crucially depends on properties of disorder
such as  preserved symmetries or the range of correlations. This has
been intensively studied using different methods during last two
decades~\cite{ludwig94,nersesyan95,horovitz02,aleiner06,%
ostrovsky06,shon98,khveshchenko07,fedorenko12}.
On the contrary, much less is known on the disorder effects close to and at the
topological semi-metal-insulator transition. Very recently, the Friedel
oscillations induced  in the
semi-Dirac fermions by a single localized impurity
have been studied in~\cite{dutreix13}. In the
present paper we address the effect of multiple randomly distributed
impurities of different types on the DOS in the vicinity of
the topological transition.  We focus on
the DOS at low energy, near the original
Dirac points, while the different problem of
diffusion of semi-Dirac fermions at higher Fermi energy will be
presented elsewhere~\cite{adroguer13}.

\section{Model} \label{sec:model}

\begin{figure}\mbox{}\hspace{18mm}
\includegraphics[width=5cm]{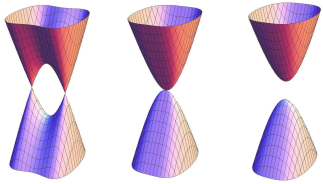}
\caption{ The energy spectrum of the clean system:
The left panel corresponds to $\Delta<0$, the right panel
to $\Delta>0$. The center panel is the semi-Dirac point $\Delta=0$. }
  \label{fig:semi-Dirac}
\end{figure}

We  consider a 2D semi-metal with valence and conduction bands
touching each other at two Dirac points. We assume that the Dirac points can
evolve in reciprocal space upon variation of the parameter $\Delta$ which
drives a topological Lifshitz transition~\cite{lifshitz60}:
for the critical value
$\Delta=0$ the two points merge as shown in fig.~\ref{fig:semi-Dirac}.
The effective Hamiltonian describing the low energy physics in the
vicinity of the transition can be derived from the tight-binding model
expanding around the merging point. It is expected to be general and
reads~\cite{montambaux09-epjb}
\begin{equation}\label{eq:Dirac2}
H_0 = \sigma_x \left(\Delta - \frac{\partial_x^2}{2m}\right)
 - i c \sigma_y \partial_y.
\end{equation}
At the critical point $\Delta=0$, Hamiltonian~(\ref{eq:Dirac2}) describes
the semi-Dirac fermions with a quadratic dispersion relation
parametrized by the mass $m$ in the
$x$-direction, and linear dispersion with velocity $c$ in the $y$-direction.
We assume that $m>0$ and set $\hbar=1$, while  $\sigma_{0}=\mathbb{I}$ and the three
$\sigma_{\mu}$, $\mu=x,y,z$ are the Pauli matrices.
The diagonalization of Hamiltonian (\ref{eq:Dirac2})
via Fourier transform gives the energy spectrum
$\varepsilon = \pm [\left(\Delta + {q_x^2}/{2m}\right)^2+ c^2 q_y^2 ]^{1/2}$.
Note that in the Hamiltonian~(\ref{eq:Dirac2}), the $x$ direction corresponding to the
real $\sigma_x$ matrix, has been identified with the direction connecting the two cones for $\Delta <0$. With this
convention, the model is clearly invariant under time reversal symmetry for any value of $\Delta$. This is indeed the
situation which arises on any tight-binding lattice model. Exchanging the $x$ and $y$ directions leads to a model
which breaks time reversal symmetry, and describes a different physics.

To incorporate in the model the presence of randomly distributed impurities
we add to Hamiltonian (\ref{eq:Dirac2}) a random perturbation
which can be parametrized in full generality by
the four disorder potentials~$V_{i}(\mathbf{r})$:
\begin{equation}\label{eq:Dirac1}
H=H_0  + \sum\limits_{i=0,x,y,z}\sigma_{i}V_{i}(\mathbf{r}).
\end{equation}
The potentials $V_{i}(\mathbf{r})$ are assumed to be random Gaussian fields
with zero mean
$\overline{V_{i}(\mathbf{r})}=0$. The
disorder strengths $\alpha_i$ parametrize the variance of the distribution according to
\begin{equation} \label{eq:Disorder}
\overline{V_{i}(\mathbf{r})V_{j}(\mathbf{r}')}=
\frac{2\pi c \alpha_{i} } {\sqrt{m}}\delta_{i j} \delta(\mathbf{r}-\mathbf{r}').
\end{equation}
All these terms have simple physical meanings:
$V_{0}(\mathbf{r})$ is a random scalar potential,
$V_{x}(\mathbf{r})$ describes local fluctuations of $\Delta$,
$V_{y}(\mathbf{r})$ is a random gauge potential acting in the $y$ - directions,
while  $V_{z}(\mathbf{r})$ plays the role of the random Dirac mass.
The system (\ref{eq:Dirac1})
resembles mathematically the Lifshitz point in anisotropic magnets
where a disordered, an ordered and a spatially modulated phase
meet~\cite{diehl00}.

\begin{figure} \mbox{}\hspace{14mm}
\includegraphics[width=6cm]{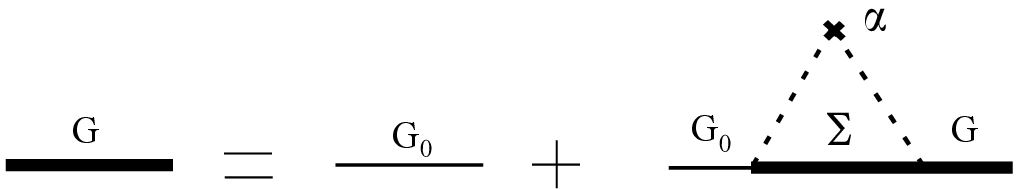}
\caption{The averaged over disorder Green function in the SCBA.}
  \label{fig:SCBA}
\end{figure}

\begin{figure}
\includegraphics[width=80mm]{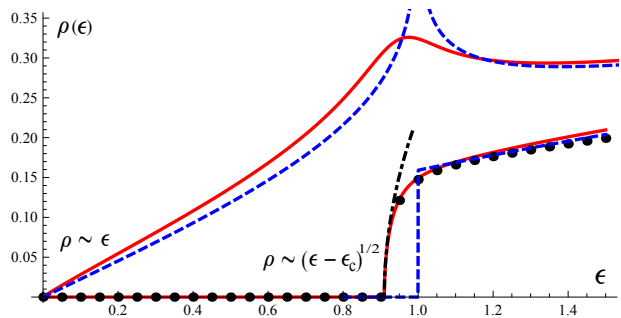}
\caption{ The DOS in units of $\sqrt{m}/c$
deep in the insulating phase for $\bar{\Delta}=1$ (bottom plot) and
in the semi-metal phase for $\bar{\Delta}=-1$ (upper plot). The
disorder strength is $\alpha=0.02$.  In the insulating phase the DOS vanishes
for $\varepsilon<\varepsilon_c\approx 0.91$. The red solid lines
are obtained using numerical
solution of eqs.~(\ref{eq-scba-alg}) and
(\ref{eq-scba-1-2-2}); blue dashed lines using the first iteration
of the latter equations that gives the shifted DOS of the clean system;
black dots using numerical solution of the simplified eq.~(\ref{eq-scba-alg-simpl}).
The black dot-dashed line is the power-law approximation (\ref{eq:fitting-1}).
  }
  \label{fig-num-sol-1}
\end{figure}

\section{The self-consistent Born approximation }

To describe the effect of disorder on the DOS at low energy, we
naturally employ the self consistent Born approximation (SCBA). The SCBA
is expected to be reliable, except possibly in the very vicinity of the
transition \cite{ostrovsky06,fukuzawa09}.
The retarded and advanced Green functions for the Hamiltonian  without disorder
(\ref{eq:Dirac2})
written in Fourier space are given by
\begin{eqnarray}
G_0^{R/A}(\varepsilon,\mathbf{q})=\frac{\varepsilon
 +\left(\Delta + \frac{q_x^2}{2m}\right)\sigma_x + c q_y \sigma_y  }{
(\varepsilon \pm 0)^2-\left(\Delta + \frac{q_x^2}{2m}\right)^2
 - c^2 q_y^2  }.
\end{eqnarray}
To calculate the corresponding Green functions averaged over disorder within
the SCBA one takes into account  the one-loop diagram contributing to
the self-energy. Replacing the bare Green function in this
diagram by the dressed one, we obtain a self-consistent equation
for the self-energy $\Sigma (\varepsilon)$ graphically depicted in
fig.~\ref{fig:SCBA}. The equation is simplified by neglecting the external momenta
dependance of  the self-energy $\Sigma (\varepsilon)$.
For convenience we introduce the matrix function
$X(\varepsilon)=\varepsilon - \Sigma (\varepsilon)$. Making an ansatz
$X(\varepsilon)= X_0(\varepsilon)\sigma_0 -X_1(\varepsilon)\sigma_x$,
where $X_i(\varepsilon)$
are scalar functions, we can rewrite the SCBA equation as
\begin{subequations}
\begin{eqnarray}
\!\!\!\!\!\!\!\!\!\!\!\! &&X_0(\varepsilon) = \varepsilon
+\frac{2\pi c \alpha } {\sqrt{m}}
\int_q
\frac{X_0(\varepsilon)  }{\Omega[X_1, q_x]^2
+c^2 q_y^2 - X_0^2(\varepsilon)}, \label{eq-scba-1}  \ \ \ \ \  \\
\!\!\!\!\!\!\!\!\!\!\!\! &&
 X_1(\varepsilon) = - \frac{2\pi c \tilde{\alpha} } {\sqrt{m}} \int_q
\frac{\Omega[X_1, q_x]
}{\Omega[X_1, q_x]^2
+c^2 q_y^2 - X_0^2(\varepsilon)}. \label{eq-scba-2} \ \ \ \ \
\end{eqnarray}
\end{subequations}
Here we have denoted $\int_q :=\int d^2 q/(2\pi)^2$
and $\Omega[X_1, q_x]:= \Delta+X_1(\varepsilon)+{q_x^2}/{(2m)}$, and also
defined the total disorder strength
$\alpha:=\alpha_0+\alpha_x+\alpha_y+\alpha_z$
and the disorder asymmetry $\tilde{\alpha}:=\alpha_0+\alpha_x-\alpha_y-\alpha_z$.
We first notice that the UV diverging integral in eq.~(\ref{eq-scba-2})
is determined by the cutoff $\Lambda$.
To simplify calculations we impose the cutoff only on wavevectors in the
$x$-direction, so that $|q_x|<\Lambda$, while we will neglect it in
the $y$-direction. Integrating over $q_x$ and $q_y$ in eq.~(\ref{eq-scba-2})
we obtain $X_1= - {\tilde{\alpha}\Lambda}/\sqrt{m} $ for large $\Lambda$ which
does not depend on $\varepsilon$.
It is convenient to introduce the renormalized  driving parameter
\begin{equation}
\bar{ \Delta}=\Delta-\frac{\tilde{\alpha}}{\sqrt{m}} \Lambda, \label{eq-shift}
\end{equation}
which accounts for the shift due to disorder asymmetry of the semi-metal to insulator transition point
from $\Delta=0$ towards $\bar{\Delta}=0$.
The shift depends on the cutoff and thus is non universal similar to what
happens at the classical Lifshitz point~\cite{diehl00}.
Solving eq.~(\ref{eq-scba-1}) for $X_0$  we find the DOS from the expression
\begin{eqnarray}
\rho(\varepsilon)= -\frac1{\pi} \mathrm{Im\ Tr} \int_q G^R(\varepsilon,\mathbf{q})
= \frac{\sqrt{m}}{\alpha c \pi^2 } \textrm{Im} X_0^R(\varepsilon). \label{eq-DOS}
\end{eqnarray}
The form of the integral over $\mathbf{q}$ in eq.~(\ref{eq-scba-1})
strongly depends on the sign of $\bar{\Delta}$, {\it i.e.} on
the direction from which we approach  the topological transition.
Hence we first discuss the effect of disorder deep
in the insulating~($\bar{\Delta}>0$) and semi-metal~($\bar{\Delta}<0$)  phases,
and finally we concentrate on the vicinity of the topological
transition~($\bar{\Delta}\approx 0$).


\subsection{The insulating phase} For $\bar{\Delta} \ge 0$,
integrating over $\mathbf{q}$ in eq.~(\ref{eq-scba-1}) we obtain
\begin{eqnarray} \label{eq-scba-alg}
X_0 = \varepsilon +  \frac{ \sqrt{2} \alpha X_0 }{\sqrt{\bar{\Delta}-X_0}}
\, K\left(2+\frac{2 \bar{\Delta}
   }{X_0-\bar{\Delta}}\right),
\end{eqnarray}
where $K(z)=\int_0^{1} {dt} [{{(1-t^2)(1-z t^2)}}]^{-1/2}$
is the complete elliptic integral of the first kind.\footnote{
Note that this definition differs from that used in \cite{montambaux09-epjb}
by changing $K(z) \to K(\sqrt{z})$.} Equation~(\ref{eq-scba-alg}) has two complex
conjugate solutions corresponding to the retarded and advanced functions.
We always find for $\bar{\Delta} >  \bar{\Delta}_c=\pi^2 \alpha^2/2$ the DOS
of an insulator: there exists an energy gap
$\varepsilon_c=\bar{\Delta} + O(\alpha)$ such that
the solution of eq.~(\ref{eq-scba-alg}) is purely real for
$0\le\varepsilon \le \varepsilon_c$, and thus, the DOS $\rho(\varepsilon)$ vanishes.
The DOS computed using eq.~(\ref{eq-DOS}) and the imaginary part of the
numerical solution of
eq.~(\ref{eq-scba-alg}) is shown in fig.~\ref{fig-num-sol-1}.
For $|\varepsilon-\varepsilon_c|\gg 0$ one can solve eq.~(\ref{eq-scba-alg})
by iterations: this provides an expansion of the solution
in powers of disorder strength.
The first iteration derived by taking $X_0 \to \varepsilon + i0^+$ in
the r.h.s of eq.~(\ref{eq-scba-alg}) gives a linear in disorder correction and
thus the DOS of the clean system found in \cite{montambaux09-epjb}
up to the shift  $\Delta \to \bar{\Delta}$.
It is also
shown in fig.~\ref{fig-num-sol-1} for comparison.
To study the solution of the SCBA equation
in the vicinity of the energy gap $\varepsilon_c$, we replace the complete
elliptic integral in eq.~(\ref{eq-scba-alg}) by its asymptotics that yields
\begin{eqnarray} \label{eq-scba-alg-simpl}
X_0 = \varepsilon +\frac{\alpha}{2} \sqrt{{X_0 }}
   \left[\ln \left(\frac{32 X_0}{X_0-\bar{\Delta}}\right)\pm i \pi  \right],
\end{eqnarray}
where "+/--" correspond to the retarded and advanced functions,
respectively.
Equation~(\ref{eq-scba-alg-simpl}) can be written as
$X_0=\varepsilon + f(X_0)$ so that the energy gap $\varepsilon_c$ and the
corresponding value $X_{0}(\varepsilon_c)=X_c$ satisfy:
$X_{c}=\varepsilon_c + f(X_{c})$ and $ f'(X_{c})=1$. In the limit of weak
disorder we obtain
\begin{eqnarray} \label{eq:Delta-c}
\varepsilon_c = \bar{\Delta} \left\{1 -
\frac{\alpha}{2 \sqrt{\bar{\Delta}}}\left[ 1 -
   \ln \left(\frac{\alpha}{64 \sqrt{\bar{\Delta}}}\right)
   \right]\right\} + O({\alpha}^2 \ln {\alpha}),
\end{eqnarray}
and the expansion of the DOS around this point
for $\varepsilon\gtrsim\varepsilon_c$:
\begin{eqnarray} \label{eq:fitting-1}
\rho (\varepsilon) = \frac{\sqrt{m}}{\alpha c \pi^2 }
 \left[ \frac{2(\varepsilon-\varepsilon_c)}{f''(X_{c})}
\right]^{1/2} + O\left((\varepsilon-\varepsilon_c)^{3/2}\right).
\end{eqnarray}
Thus, the DOS of the disordered system in the insulating phase vanishes as a power
law with the exponent $\frac12$ in contrast to the jump in
the DOS of the clean
system (see the bottom plot in fig.~\ref{fig-num-sol-1}).

\subsection{The semi-metallic phase}
In the case $\bar{\Delta}<0$, the
integration over $\mathbf{q}$  in eq.~(\ref{eq-scba-1})  leads to the equation
\begin{equation} \label{eq-scba-1-2-2}
X_0 = \varepsilon +
   \frac{ \sqrt{2} \alpha X_0 }{\sqrt{\bar{\Delta}+X_0}}
   \left[
    K\left(\frac{2X_0}{\bar{\Delta}+X_0}\right) +
    2 i K\left(\frac{\bar{\Delta}-X_0}{\bar{\Delta}+X_0}\right)
 \right].
\end{equation}
The imaginary part of the numerical solution of eq.~(\ref{eq-scba-1-2-2})
gives the DOS which is  shown in the upper plot of fig.~\ref{fig-num-sol-1}.
Iterating eq.~(\ref{eq-scba-1-2-2}) one can obtain the solution
for $|\varepsilon-\bar{\Delta}|\gg 0$. The DOS of the clean system
found in \cite{montambaux09-epjb}
(up to the shift (\ref{eq-shift})) is recovered by taking the limit
$X_0 \to \varepsilon + i0^+$ in the r.h.s of eq.~(\ref{eq-scba-1-2-2}) and
substituting the solution to eq.~(\ref{eq-DOS}). This DOS
diverges  logarithmically at $\varepsilon=|\bar{\Delta}|$
while the DOS of the disordered system is continuous and smooth
(see the upper plot in fig.~\ref{fig-num-sol-1}).

\begin{figure} \mbox{}\hspace{7mm}
\includegraphics[width=75mm]{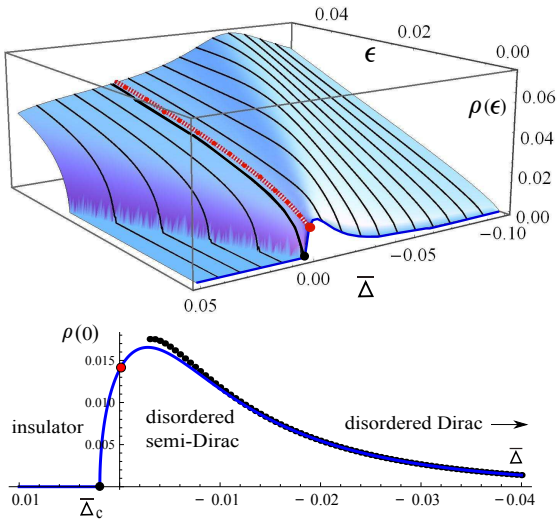}
\caption{ The DOS in units of $\sqrt{m}/c$ close to the transition
as a function of $\bar{\Delta}$ (upper panel at arbitrary energy, bottom panel at zero
energy). The disorder strength is $\alpha=0.02$.
The red dashed line is the DOS at  $\bar{\Delta}=0$ given by eq.~(\ref{eq-rho-tr}).
The DOS at the critical point $\bar{\Delta}_c\approx 0.002$ is given by
eq.~(\ref{eq-dos-critical}). Dotted line in the bottom panel is given
by eq.~(\ref{eq-dos-zero-energy}).
  }
  \label{fig-num-sol-3}
\end{figure}

\subsection{Topological transition}
Let us now focus on the region of small $\overline{\Delta}$.
The DOS computed numerically using the SCBA
eqs.~(\ref{eq-scba-alg}) and (\ref{eq-scba-1-2-2})
is shown in fig.~\ref{fig-num-sol-3}  as a function of $\bar{\Delta}$
in the vicinity of the transition $\bar{\Delta}=0$.
One can see that in the presence of disorder there is a region around the
transition point $\bar{\Delta}=0$  characterized by a finite DOS
at zero energy, $\rho(\varepsilon=0)>0$. The DOS at $\varepsilon=0$ strictly
vanishes for $\bar{\Delta}>\bar{\Delta}_c=\pi^2 \alpha^2/2$.
Thus, in the presence of disorder the transition (semi-Dirac) point
between  the Dirac semi-metal and the insulator is replaced by a
new intermediate regime which we call
the disordered semi-Dirac regime.
The disordered Dirac phase smoothly evolves into this intermediate
regime, while the SCBA suggests  a sharp continuous transition
at $\bar{\Delta}=\bar{\Delta}_c$ between this disordered semi-Dirac phase
and the insulating phase. Exactly at the transition we find an algebraic
behavior of the DOS
\begin{equation} \label{eq-dos-critical}
\rho\left(\varepsilon;\bar{\Delta}=\bar{\Delta}_c\right)=
\left(\frac34\right)^{1/6}  \frac{\sqrt{m}}{ c \pi }
( \pi \alpha )^{1/3} {\varepsilon }^{1/3},
\end{equation}
which is displayed in fig.~\ref{fig-num-sol-3}.
 The extrapolation towards the disordered semi-metallic regime is obtained
for large negative $\bar{\Delta}<0$: in this regime we recover a DOS at
zero energy decaying as
\begin{eqnarray} \label{eq-dos-zero-energy}
\rho(0)= \frac{8|\bar{\Delta}|\sqrt{m}}{\alpha c \pi^2} e^{- \sqrt{|\bar{\Delta}|}
 /(\alpha\sqrt{2})}.
\end{eqnarray}
Hence, instead of a transition the system undergoes a crossover to
disordered Dirac fermions. Indeed, this behavior is fully consistent
with the SCBA picture for the disordered
Dirac fermions. For graphene it predicts a finite DOS
$\rho(\varepsilon)= 2\Gamma_0/(\pi^2 v_0^2 \alpha_D )$
for $\varepsilon \ll \Gamma_0$ where $\Gamma_0=\Delta_D e^{-1/\alpha_D}$ is
the energy scale generated by disorder.
Here $\Delta_D$ is the bandwidth, $v_0$ - the Fermi velocity and $\alpha_D$ - the
dimensionless strength of disorder~\cite{ostrovsky06,shon98}.
This would imply a finite DOS at zero energy for all
considered types of disorder. However, this contradicts the results obtained for
random gauge and random mass disorder using more reliable methods.
For instance, for random
gauge disorder one expects $\rho(\varepsilon)=\varepsilon^{2/z-1}$ with
$z=1+\alpha_D$ in weak disorder case ($\alpha_D<2$)~\cite{ludwig94} and
$z = (8\alpha_D)^{1/2} -1$ in strong disorder
case~($\alpha_D>2$)~\cite{horovitz02}. Similarly, with random mass disorder, the DOS
remains linear in $\varepsilon$ up to logarithmic corrections~\cite{fedorenko12}.
Only for scalar potential disorder the DOS saturates at a
finite value in the vicinity of the Dirac point~\cite{ostrovsky06}.
It was argued in~\cite{nersesyan95}  that the failure of SCBA in
the vicinity of the Dirac point for $\varepsilon< \Gamma_0$
is due to importance of the diagrams with crossed disorder lines, neglected within
the SCBA which takes into account only the non-crossed ones. Another possible reason
for the failure of the SCBA is   the divergence of the fermion wavelengths
at the Dirac point  rendering the SCBA uncontrollable, i.e.
without a small parameter. Nevertheless one can
still rely on the SCBA for energies $\varepsilon\gg \Gamma_0$.
The question of reliability of the SCBA in the vicinity of the topological
transition is even more subtle.

\subsection{Disordered Semi-Dirac}
It is interesting to note that the SCBA equations can be analytically
solved exactly at the point $\bar{\Delta}=0$, inside
the ``disordered semi-Dirac'' regime. Indeed using that
$K(2+0^+/z)= C(1-\mathrm{sign}[\mathrm{Im}\, z]i)/ \sqrt{2}$
where
$ C = { \Gamma \left({1}/{4}\right)^2}/{(4 \sqrt{\pi} )}$
one can rewrite
eq.~(\ref{eq-scba-alg}) as
$X_0=\varepsilon +(1\mp i) C {\alpha}  \sqrt{-X_0}$ ,
where "-/+" corresponds to the retarded/advanced Green's function.
We obtain the DOS at the point $\bar{\Delta}=0$
\begin{eqnarray}
\rho (\varepsilon)= \frac{\sqrt{m} C^2 \alpha}{c \pi^2}
\left[
1+ \frac1{\sqrt{2}} \sqrt{1+\sqrt{1+ \frac{4 \varepsilon^2}{C^4 {\alpha}^4} }}
\right], \label{eq-rho-tr}
\end{eqnarray}
such that $\rho (0)= 2{\sqrt{m} C^2 \alpha}/{(c \pi^2)}$.
Note that the expansion of eq.~(\ref{eq-rho-tr}) in powers of the disorder
strength $\alpha$  is in fact an expansion
in ${\alpha}/{\sqrt{\varepsilon}}$.
Its truncation to a finite order in $\alpha$
appears to be pathological for $\varepsilon \ll \alpha^2$.
We interpret this as a sign of the probable
breakdown of the SCBA for a small enough energy. In this regime, a more
refined technique is required~\cite{aleiner06}.

Before we consider a weak disorder renormalization group (RG) approach to the
topological transition let us first discuss the consequences of our SCBA results
for the behavior of specific heat and spin magnetic
susceptibility~\cite{montambaux09-epjb,banerjee12}.
Assuming that the chemical potential is temperature independent and fixed at
zero energy the specific heat can be written as eq.~(18) in~\cite{montambaux09-epjb}.
In the clean system the low $T$ specific heat interpolates from a $T^2$ behavior
far in the semi-metal phase to an activated behavior in the insulating phase.
Exactly at the transition it exhibits a $T^{3/2}$ behavior. In the presence
of disorder  the low $T$ specific heat at the critical point should behave as
$T^{4/3}$. The magnetic susceptibility $\chi$ is related to the
DOS by eq.~(10.11) in~\cite{ziman-book}. In the
clean system, $\chi$ is activated in the insulator, $\chi \sim T$
in the semi-metal phase and $\chi \sim T^{1/2}$ at the transition. With disorder,
the scaling behavior of the susceptibility at the transition is modified  to $\chi \sim T^{1/3}$.

\section{Weak disorder renormalization  group}
We now study whether the  critical point found using the SCBA  is accessible
within a weak disorder RG approach.
The starting point is the generating functional for the Green functions
$\mathcal{Z}=\int D\bar{\psi} D\psi e^{-\mathcal{S}}$, where
$\mathcal{S}$ is the effective action at a given energy (Matsubara frequency)
$\varepsilon$. The effective action can be written in terms of two fermionic
(anticommuting) fields ${\psi}(\mathbf{r})$ and $\bar{\psi}(\mathbf{r})$ which are
two completely  independent Grassman variables.
For generality  we consider a class of $d$ dimensional models
in which the bare dispersion  is quadratic  in  $l$ direction and
linear in $d-l$ remaining directions.
For instance, the case $(d=2$, $l=1)$ describes the semi-Dirac fermions,
$(d=3$, $l=0)$ - Weyl fermions~\cite{delplace12}, and
$(d=3$, $l=1\ \mathrm{or}\ 2)$ - semi-Weyl fermions,\textit{ e.g.}
similar to that found in  ${}^3\mathrm{He-A}$~\cite{volovik01}.

\subsection{Scaling dimensions}
The standard way to generalize the model to $d$ dimensions
is to introduce the generators of the $d$-dimensional
Clifford algebra.
In even dimensions these generators can be represented by the matrices
$\gamma_i$\ $(i=1,...,d)$ of dimension $2^{d/2}$. Generally one also defines
$\gamma_{d+1} \equiv \gamma_{S} = i^{-d/2}  \gamma_1 ...  \gamma_d$.
For the case of odd dimensions one constructs the matrices
$\gamma_i$\ $(i=1,...,d-1)$ similarly to the case of even dimension and
adds
$\gamma_{d} \equiv \gamma_{S} = i^{-(d-1)/2}  \gamma_1 ...  \gamma_{d-1}$.
These matrices  satisfy the anticommutation relations:
$\gamma_i \gamma_j + \gamma_j \gamma_i = 2 \delta_{ij} \gamma_0$,
$i,j=1,...,[[d+1]]$,
where $\gamma_0=\mathbb{I}$ is the identity matrix and $[[d+1]]$ stands for $d+1$ if $d$ is even,
and  $d$ if it is odd.
The matrices $\gamma_i$ and $\gamma_S$ replace the Pauli matrices $\sigma_i$ ($i=x,y$)
and $\sigma_z$ in $d$ dimensions.
We define the two matrix vectors $\gamma_{\parallel}:=(\gamma_1,...,\gamma_{l})$ and
 $\gamma_{\bot}:=(\gamma_{l+1},...,\gamma_d)$ and
replace in Hamiltonian (\ref{eq:Dirac2}) $\partial_x\to \partial_{\|}$  which acts in
$l$-dimensional
subspace and  $\partial_y \to \partial_{\bot}$ which acts in
$(d-l)$-dimensional subspace.
Introducing $n$ copies of the original system and using the replica trick
to average over disorder we can write the effective replicated
action as
\begin{eqnarray} \label{eq:action-1}
&& \!\!\!\!\!\!\!\!\!\!\!\! \mathcal{S} = \int d^d \mathbf{r}\left
\{ \sum\limits_{a=1}^n
 \bar{\psi}_{a}\left[ \gamma_{\parallel} \left(\Delta - \frac{\partial_{\|}^2}{2m}\right)
 - i c \gamma_{\bot} \partial_{\bot} - i \varepsilon \right]\psi_{a}
\nonumber \right.\\
&& \!\!\!\!\!\!\!\!\!\!\!\!
 -  \frac{\pi c}{\sqrt{m}} \sum\limits_{a,b=1}^n \sum\limits_{j=0}^{[[d+1]]} \alpha_{j}
 [\bar{\psi}_{a}(\mathbf{r})\gamma_{j} \psi_{a}(\mathbf{r})]
[\bar{\psi}_{b}(\mathbf{r})\gamma_{j} \psi_{b}(\mathbf{r})] \Big\}, \
\end{eqnarray}
where $d^d \mathbf{r}:=  d^{l} \mathbf{r}_{\|}\,  d^{d-l} \mathbf{r}_{\bot}$.
The  properties of the initial disordered model are  obtained in  the limit $n\to 0$.
Dimensional analysis of the
action~(\ref{eq:action-1}) yields the bare dimensions:
$[\mathbf{r}_{\|}]= \mu^{-1/2}$, $ [\mathbf{r}_{\bot}]=\mu^{-1}$,
$ [\psi]= [\bar{\psi}] =  \mu^{(2d-l-2)/4}$ and $[\alpha_j] =  \mu^{\delta}$,
where  $\mu$ is an arbitrary momentum scale and
$\delta=2+{l}/2 -d$.
Thus the upper critical dimension of the model~(\ref{eq:action-1}) is
$d_c=2+l/2$ and disorder is
(i) relevant for $\delta>0$, e.g. for 2D semi-Dirac fermions with $\delta=1/2$;
(ii) marginally relevant  for 2D Dirac fermions with $\delta=0$;
(iii) irrelevant  for 3D Weyl ($\delta=-1$) and semi-Weyl fermions ($\delta=-1/2$).

\subsection{One-loop RG with restricted disorder} 
\begin{figure} \mbox{}\hspace{14mm}
\includegraphics[width=60mm]{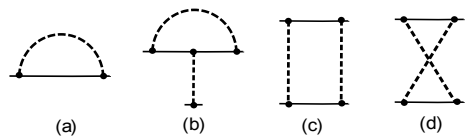}
\caption{The one-loop diagrams contributing to the propagator (a) and
disorder renormalization  (b)-(d). }
  \label{fig:diagrams}
\end{figure}

We focus here on the case of 2D
semi-Dirac fermions.
To renormalize the model (\ref{eq:action-1}) one has to eliminate
UV divergences in the loop expansion by introducing counter terms.
The one-loop counterterms to the propagator and disorder are given by the diagrams
shown in fig.~\ref{fig:diagrams}. To one-loop order the $m$ and $c$ do not get
corrected while $\Delta$ gets an additive shift similar to that in
eq.~(\ref{eq-shift}). We use dimensional regularization in which
this shift vanishes \cite{diehl00}.
The diagrams (c) and (d) in fig.~\ref{fig:diagrams} can generate extra
terms with new algebraic structures like
$\gamma_i \gamma_j \gamma_k \otimes \gamma_i\gamma_j\gamma_k$ and so on.
In $d=d_c-\delta$ dimensions they have to be treated  as independent new
interactions reflecting the fact that within dimensional regularization
this theory is not multiplicatively renormalizable \cite{bennett99}.
For the Dirac fermions
this problem shows up to orders higher than one-loop and  can be solved, for example,
using a cutoff in strictly 2D
so that the extra interactions do not arise. In contrast, for the semi-Dirac
fermions the extra interactions infinitely proliferate in the renormalization
already to the one-loop order. This cannot be cured by using a
different regularization scheme since the upper critical dimension $\frac52$ is
non integer.
To avoid this obstacle we restrict ourselves to the
presence of a single type of disorder. It is easy to see that the diagrams
(c) and (d) in fig.~\ref{fig:diagrams} with any type of disorder always contribute
to $\alpha_1\equiv \alpha_x$. Thus we can derive a closed one-loop flow equation
 by considering a model with  only the $\alpha_x$ type of disorder.
 The corresponding RG equations read
\begin{equation}
-\mu \partial_\mu \ln \varepsilon = 1+\alpha_x,   \  \  \  \
-\mu \partial_\mu \alpha_x = \delta\, \alpha_x -\alpha_x^2, \label{eq-flow-alpha}
\end{equation}
where we have included the value of the one-loop integral in redefinition
of $\alpha_x$.
Equations~(\ref{eq-flow-alpha}) possess a fixed point $\alpha_x^*=\delta$.
Integrating them up to the scale corresponding to $\varepsilon \simeq 1$,
we find $\varepsilon=\mu^{z}$ with
$z=1+\alpha_x^*=1 +\delta + O(\delta^2)$.
The DOS behaves as $\rho\sim \langle\bar{\psi} \psi\rangle$, which leads to its scaling as
$\rho\sim \mu^{3/2-z}$
or equivalently in terms of energy as
\begin{equation}
\rho(\varepsilon) \sim \varepsilon^{(3/2-z)/z}=
\varepsilon^{(1/2-\delta)/(1+\delta)+O(\delta^2)}.
\end{equation}
Hence we find that for $\delta=\frac12$ the DOS exponent vanishes to
one loop order suggesting a constant DOS at zero energy when
only the $\alpha_{x}$ disorder is present. This result improves
upon the critical behavior extracted from the SCBA analysis which was shown to be
invalid at low energy. Whether the exponent vanishing is accidental or holds
beyond the one-loop order  remains an open question.

\subsection{Full one-loop RG analysis}
 To analyze the scaling behavior of the full disordered 2D semi-Dirac fermions, we find it more technically
 tractable to consider an alternative generalized model given by the effective action
\begin{eqnarray} \label{eq:action-2}
&& \!\!\!\!\!\!\!\! \mathcal{S} = \int d^d \mathbf{r}\left \{ \sum\limits_{a=1}^n
 \bar{\psi}_{a}\left[ \sigma_x \left(\Delta - \frac{\nabla_{\|}^2}{2m}\right)
 - i c \sigma_y \partial_y - i \varepsilon  \right]\psi_a
\nonumber \right. \\
&& \!\!\!\!
 -  \frac{\pi c}{\sqrt{m}} \sum\limits_{a,b=1}^n \sum\limits_{j=0,x,y,z} \alpha_{j}
 [\bar{\psi}_{a}(\mathbf{r})\sigma_{j} \psi_{a}(\mathbf{r})]
[\bar{\psi}_{b}(\mathbf{r})\sigma_{j} \psi_{b}(\mathbf{r})] \Big\}. \nonumber
\end{eqnarray}
where $d^d \mathbf{r}:=  d^{d-1} \mathbf{r}_{\|}\,  d \mathbf{r}_{\bot}$.
The advantage of this model is that it involves no infinite Clifford algebra.
Dimensional analysis of this model suggests that its upper critical dimension
is $d_c=3$. We can construct a perturbative in disorder RG around this dimension
using expansion in a small parameter $\delta=(3-d)/2$ similar to that done for the
model~(\ref{eq:action-1}). The corrections to energy and disorder strengths are
given by the same diagrams shown in fig.~(\ref{fig:diagrams}).
The one-loop RG equations read
\begin{eqnarray}
\!\!\!\!\!\! && -\mu \partial_\mu \ln \varepsilon = 1+\alpha_0+\alpha_x
  +\alpha_y+\alpha_z,  \ \ \ \ \ \ \ \ \nonumber \\
\!\!\!\!\!\! && -\mu \partial_\mu \alpha_0 = \delta\, \alpha_0 +
   4 \alpha_0 (\alpha_0+2\alpha_x+\alpha_y+\alpha_z)
    + 4 \alpha_x \alpha_z, \ \ \ \ \ \ \ \ \nonumber \\
\!\!\!\!\!\!&& -\mu \partial_\mu \alpha_x = \delta\, \alpha_x
  + 2 \left(\alpha_0^2+\alpha_x^2+\alpha_y^2+\alpha_z^2\right)
    +4 \alpha_0 \alpha_z,  \ \ \ \ \ \ \ \ \nonumber\\
\!\!\!\!\!\!&& -\mu \partial_\mu \alpha_y = \delta\, \alpha_y
   + 4 \alpha_x \alpha_y, \ \ \ \ \ \ \ \ \nonumber \\
\!\!\!\!\!\! &&  -\mu \partial_\mu \alpha_z = \delta\, \alpha_z
 -4 \alpha_z (\alpha_0-2\alpha_x-\alpha_y+\alpha_z)
     + 4 \alpha_0 \alpha_x. \ \ \ \ \ \ \ \ \nonumber
\end{eqnarray}
Unfortunately, the solutions of these RG equations always flow towards a strong
disorder regime which describes Anderson localization,
but is not within the reach of a weak disorder RG analysis. In this generalized
model, we do no find a critical perturbative fixed point,
as opposed to the model with only $\alpha_{x}$ disorder.

\section{Conclusions}

We have studied the effect of disorder  on  the topological merging transition
from a Dirac semi-metal to a band insulator in 2D systems. Using the SCBA we
have found that disorder smears out the transition and an intermediate disordered
semi-Dirac regime emerges.
This phase is separated from the insulating phase by a sharp continuous transition
while there is a crossover between this phase and disordered Dirac fermions regime.
Exactly at the transition the SCBA suggests a power-law behavior of DOS with
a nontrivial exponent $\frac13$.
We have also shown that  the critical point can be perturbatively described
using a weak disorder RG when only random  fluctuations of $\Delta$
are present but is unfortunately inaccessible for the most general model. It would
also be  interesting to study the effect of disorder on other types of merging
transitions~\cite{volovik07}.

\acknowledgments

We would like to thank G. Montambaux for inspiring and useful discussions.
This work has been supported  by the ANR project 2010-BLANC-041902 (IsoTop).

\end{document}